\documentclass[%
 reprint,
showpacs,preprintnumbers,
showkeys
 amsmath,amssymb,
 aps,
]{revtex4-1}

\usepackage{graphics}
\usepackage{dcolumn}
\usepackage{bm}

\usepackage[usenames, dvipsnames]{color}
\usepackage{epsfig}
\usepackage{amsmath}

\graphicspath{{figs/}}

\usepackage[mathlines]{lineno}
\usepackage{soul}

\def\mtant{{\mbox{Ta}_2\mbox{O}_5}}
\def\tant{$\mtant$}
\def\sil{$\mbox{SiO}_2$}
\def\ttan{$\mbox{TiO}_2$}
\def\Cfold{C_{\rm fold}}

\newcommand{\SI}[2]{\ensuremath{#1\,{\rm #2}}}

\def\rtHz{\sqrt{\rm Hz}}
\def\mrtHz{{\rm m}/\rtHz}
\def\HzrtHz{{\rm Hz}/\rtHz}
\def\frmrtHz{\frac{\rm m}{\rtHz}}

\def\Pabs{P_{\rm abs}}

\definecolor{spring}{rgb}{0.7,0.9,0.7}
\definecolor{brick}{rgb}{0.7,0.2,0.1}
\definecolor{redHL}{rgb}{1.0,0.5,0.5}

\begin{document}


\title{Audio-band Coating Thermal Noise Measurement for Advanced LIGO 
with a Multi-mode Optical Resonator}

\author{S. Gras}%
\author{H. Yu}%
\author{W. Yam}
\author{D. Martynov}
\author{M. Evans}

\affiliation{Massachusetts Institute of Technology, 185 Albany St. NW22-295, 02139 MA, USA}
 \email{sgras@ligo.mit.edu}


\date{\today}

\begin{abstract}
In modern high precision optical instruments, such as in gravitational wave detectors
 or frequency references, thermally induced fluctuations in the reflective coatings
 can be a limiting noise source. 
This noise, known as coating thermal noise, can be reduced by choosing materials with low mechanical loss. Examination of new materials becomes a necessity in order to further minimize the
 coating thermal noise and thus improve sensitivity of next generation instruments.
We present a novel approach to directly measure coating thermal noise using a high finesse folded cavity in 
 which multiple Hermite-Gaussian modes co-resonate. 
This method is used to probe surface fluctuations on the order $10^{-17} \mrtHz$
 in the frequency range \SI{30-400}{Hz}.
We  applied this technique to measure thermal noise and loss angle
 of the coating used in Advanced LIGO.
\end{abstract}

\maketitle

\section{Introduction}

Dielectric coatings used in high precision optical instruments consist of alternating
 layers of materials with low and high index of refraction.
Thermal noise in these coatings arises from mechanical dissipation in the coating materials described by the fluctuation dissipation theorem.
This noise limits the sensitivity of the current gravitational wave detectors~\cite{0264-9381-24-2-008, PhysRevD.78.102003, PhysRevLett.116.061102}, of the best frequency references~\cite{Ludlow:07}, and of macroscopic quantum measurements~\cite{1367-2630-11-7-073032, Poot2012273}. 

Further improvement of these instruments calls for reduction of the coating thermal noise. The materials presently in use belong to the class of amorphous glassy oxides including SiO$_2$, Ta$_2$O$_5$, ZrO$_2$, Nb$_2$O$_5$, HfO$_2$ and Al$_2$O$_3$.
The search for the new high reflectivity surfaces with low mechanical loss explores a wide range of possibilities: from new amorphous coatings produced with conventional ion beam sputtering techniques \cite{0264-9381-27-8-084030, PhysRevD.91.042002}, to crystal coatings \cite{nphoton.2013.174} and grating reflectors \cite{PhysRevD.88.042001}.

The loss angle of new coating materials is most frequently obtained based on the measurement of the mechanical quality factor. The techniques presented in the literature include, among others,
 suspended disks~\cite{crooks_blades, harry_blades},
 clamped cantilevers~\cite{pierro_cantilevers},
 and the gentle nodal suspension~\cite{cesarini_nodal}.
The level of coating thermal noise is then calculated from the measured parameters,
 such as mechanical loss angles, Poisson ratio, and Young's modulus.
However, due to uncertainties in the multilayer parameters a robust experimental setup
 is necessary to directly measure coating thermal noise of a particular sample.

Such a measurement is complicated by multiple noise sources such as table vibrations, laser frequency and amplitude noise, and various readout noises.
In the past, direct measurements of the coating thermal noise have been
 accomplished using suspended free-space Fabry-Perot cavities~\cite{numata_ctn, black_ctn}.
Seismic motion limits the sensitivity of these experiments below 100\,Hz. On the other hand, fixed-spacer cavities with optically contacting mirrors were recently developed to observe coating thermal noise below 100\,Hz~\cite{tara_ctn}. However, the readout of this experiment is located in transmission of the cavities. This sets an upper limit on the reflectivity of the measured sample.

This paper describes a novel technique for the direct observation of the coating thermal noise which
 uses only one free-space Fabry-Perot cavity, and in which there is no upper limit on the sample reflectivity.
Multiple transverse electromagnetic modes (TEM) co-resonate in the cavity: 00, 02 and 20.
These modes have orthogonal spacial profiles, and probe different areas of the sample coating,
 while other displacement noises of the cavity are common to all resonating modes.
Coating thermal noise is extracted from the frequency difference between the two higher order modes.

In Sec.~\ref{ctn} we describe analytical calculations of the coating thermal noise for the fundamental and higher
 order modes in the linear and folded cavities. Sec. \ref{exp} describes our experimental setup. We have used it to measure the coating thermal noise of an Advanced LIGO~\cite{design_aligo} witness sample.
In Sec. \ref{res} we discuss the sensitivity of our experiment,
 measured coating thermal noise of Advanced LIGO sample and the estimation of \ttan:\tant\  loss angle.

\section{Coating thermal noise}
\label{ctn}

The reflectivity of an optical coating scales with the number of coating layers
 and for typical coating materials, a transmission of a few parts-per-million can be achieved with a coating roughly 10 optical wavelengths thick.
While excellent optical properties are available for large ($> \SI{10}{cm}$) optics with ion beam sputtered
 coatings, the metal oxides are mechanically much lossier than the fused silica or silicon substrates.
This loss makes coatings a dominant source of thermal noise \cite{nphoton.2013.174}.

The fluctuation-dissipation theorem \cite{0034-4885-29-1-306} connects the properties of an observable, in our case the displacement of the mirror surface,
 with the conversion of mechanical energy to heat (i.e., dissipation in the mirror coating).
The single sided power spectral density (PSD) of the observable is given by
\begin{equation}\label{eqn:ctnW}
S_{x}(f) = \frac{2k_BT}{\pi^2 f^2}\frac{W_{\rm diss}(f)}{F_{0}^2},
\end{equation}\\
where $T$ is the temperature,  $k_{B}$ the Boltzmann constant, and $W_{\rm diss}$ is the time averaged dissipated power in the coating when subjected to a sinusoidally varying force $F(t) = F_{0} \cos{2\pi ft}$ \cite{PhysRevD.57.659}.
Though there are a variety of dissipation mechanisms in the coating which can cause the observable to fluctuate \cite{PhysRevD.78.102003, Braginsky19991},
in this paper we focus on the dominant dissipative mechanism, mechanical loss of the coating materials.
Though this is a subset of all coating thermal noises,
 we will refer to the noise related to this dissipation mechanism as coating thermal noise (CTN).

The features of coating thermal noise can be clearly examined if we consider a simplified model
 of the coating as a single lossy layer of thickness $d$.
The power dissipation in a single layer can be written as
\begin{equation}\label{eqn:W}
W_{\rm diss}(f) = \frac{2F^2_{0} d (1+\sigma)(1-2\sigma)\phi}{\omega^2_{0} Y}\times f,
\end{equation}
where $\sigma$ and $Y$ are the Poisson ratio and Young's modulus,
 $\omega_{0}$ is the beam waist size, and $\phi$ is the mechanical loss angle.
By combining Eqns.~\ref{eqn:ctnW} and \ref{eqn:W} we can see a $1/f$
 dependence of the PSD of CTN, assuming that $\phi$ and other
 mechanical properties are frequency independent \cite{PhysRevD.91.042002}. 

\begin{figure}
 \centering
\includegraphics[scale=0.32, bb= 0 0 774 262]{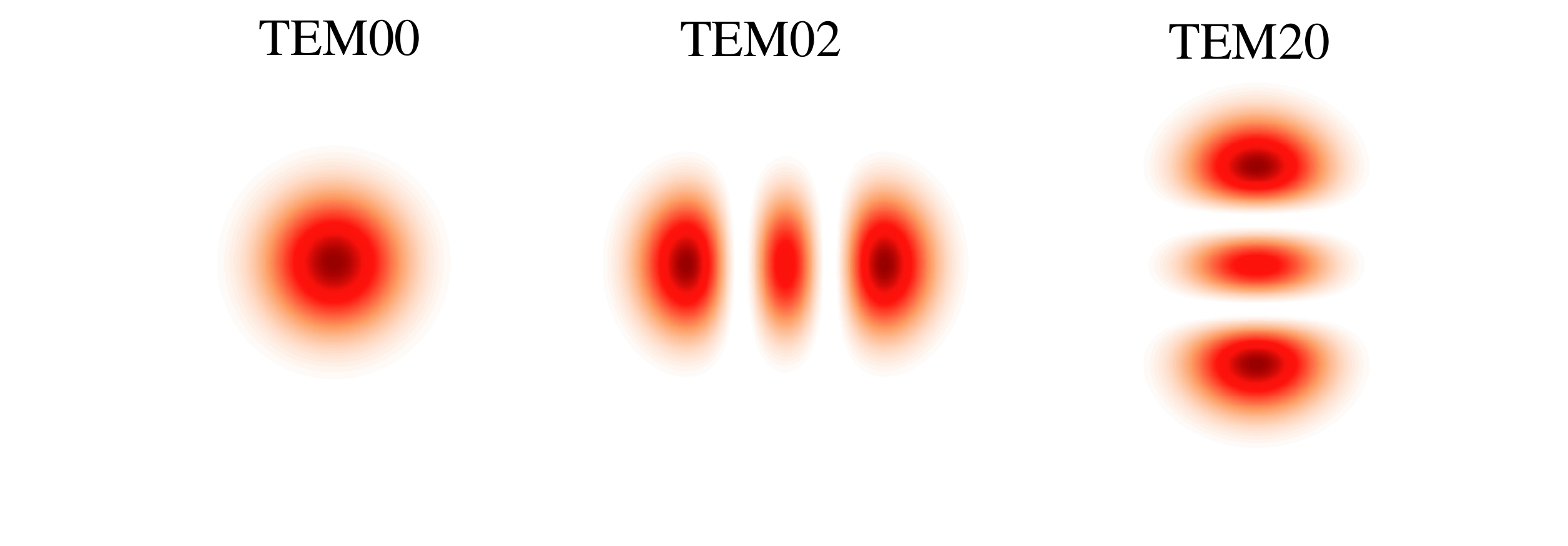}
\caption{\label{fig:tem}At the heart of this experimental work is a high-finesse optical cavity which resonates three distinct fields: a horizontally polarized Gaussian TEM00 mode, and two vertically polarized Hermite-Gaussian modes, TEM02 and TEM20 \cite{book:Lasers}. The primary advantage of using multiple resonant fields in a high-finesse cavity is that all of these fields share the same sensitivity to changes in cavity length and laser frequency. On the other hand, each mode samples a different part of the coating and thus the coating thermal noise seen by each of the resonant modes is largely independent. Since this experiment measures the difference between the resonant frequencies of the TEM02 and TEM20 modes, ideally all cavity length noises cancel leaving only the desired coating thermal noise.} 
\end{figure}

While direct measurements of the coating thermal noise are associated with a particular set of parameters,
 i.e. beam size, beam  spatial profile, and cavity geometry,
  it is often necessary to predict the level of the coating thermal noise for different parameters.
 In particular, our experiment measures the thermal noises sensed by TEM02 and TEM20 modes,
  shown in Fig.~\ref{fig:tem} and referred to as $N_{\rm CTN}$.
On the other hand, the lowest-order transverse mode TEM00 is commonly used in optical experiments such as gravitational wave observatories for which the coating thermal noise equals to $N_{\rm CTN}^{00}$. 
In order to estimate $N_{\rm CTN}^{00}$ from $N_{\rm CTN}$, we employ the correction factor $C$:
\begin{equation}\label{eqn:sctn1}
	N_{\rm CTN}^{00}=\sqrt{C}N_{\rm CTN}.
\end{equation}  
A detailed description of our computation of this correction factor can be found in Sec.~\ref{sec:C}. 

\section{Experimental Setup}\label{exp}

The key component of the measurement technique presented in this article is the folded cavity with three co-resonant optical modes: TEM02, TEM20, and TEM00.
The TEM02 and TEM20 second order transverse modes, collectively referred to as ``higher order modes'' (HOMs), are shown in Fig.~\ref{fig:tem}.
The thermal noise sensed by the TEM02 mode differs from
 the noise sensed by the TEM20 mode since these modes sample different areas of the coating.
Both resonant modes have, on the other hand, nearly identical response to the common mode noises such as laser frequency noise, cavity length noise, and mirror substrate thermal noise.
The TEM00 mode, which also shares the same sensitivity to the laser frequency and cavity length, is used to suppress these common noises (see Section \ref{sec:layout}). The primary output of the experiment is the difference between the resonant frequencies of TEM02 and TEM20.

The TEM02 and TEM20 modes are chosen for a number of reasons.
First, since they are even order modes, coupling into these modes has
 no first order sensitivity to the alignment of the cavity relative to the input beam~\cite{Anderson:84}.
Secondly, even order modes of the optical cavity can be excited by the input beam in the fundamental mode. No special optics are required to achieve 12\% of the power coupling.
Lastly, modes of the same order are required to maintain a small separation in their resonant frequencies and keep high common mode rejection to the cavity noises (see Sec.~\ref{res:vibration}).

The remainder of this section describes the parameters of the experiment such as
 geometry of the optical cavity, input and output optics, readout technique and feedback control loops.

\subsection{Optical cavity}

\def\fFSR{f_{\rm FSR}}
\def\fTMS{f_{\rm TMS}}

\begin{figure}
\includegraphics[scale=0.45]{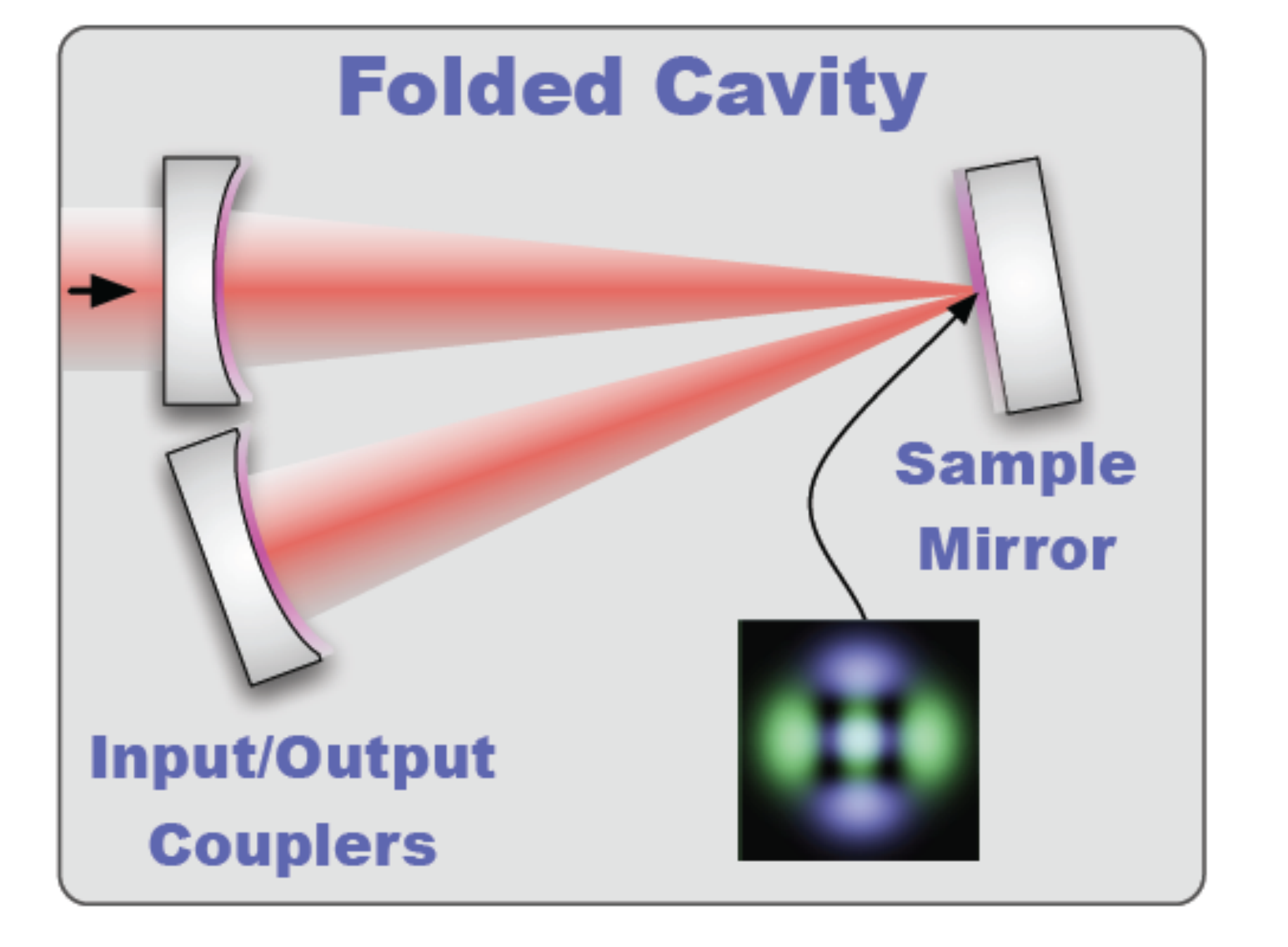}
 \caption{\label{fig:cavity}While the multi-mode approach can be applied to any optical cavity, this experiment benefits greatly from a folded geometry. Among the numerous advantages of this geometry are the use of a flat high-reflectivity sample mirror, and the ability to change the size of the beam on the sample mirror without changing the optical modes resonant in the cavity. The inset image shows the TEM20 and TEM02 modes, highlighting the fact that they overlap only in a small central area and otherwise sample distinct regions of the coating.} 
\end{figure}

The experiment uses a 3-mirror folded cavity, with the sample to be measured as the folding mirror (see Fig.~\ref{fig:cavity}). The cavity is located in vacuum at a pressure of $10^{-5}$\,Torr and at room temperature.
This folded configuration allows us to test high reflectivity coatings, and, since the sample mirror is flat, we can use the witness flats commonly included in the coating fabrication process of large optics.
This configuration also allows us to change the size of the beam on the sample mirror by
 changing the location of the sample mirror in the cavity (without changing the cavity length),
 thereby enabling an exploration of the scaling of coating thermal noise with beam size.

\subsubsection{Geometrical parameters}\label{exp:cav:geom}

We chose a folded cavity of nominal length $L = \SI{9.8}{cm}$ and input and output couplers
 with radius of curvature $R = \SI{5.0}{cm}$.
These parameters imply that a waist size $\omega_0$,
 free spectral range $\fFSR$, and the transverse mode spacing
 $\fTMS$ \cite{book:Lasers} of
\begin{equation}
\begin{split}
	& \omega_{0} = \sqrt{\frac{\lambda\sqrt{L\epsilon}}{\pi}}  \simeq \SI{49}{\mu m} \\
	& \fFSR = \frac{c}{2L} \simeq \SI{1.53}{GHz} \\
	& \fTMS = \frac{c}{\pi L}\sqrt{\frac{\epsilon}{R}} \simeq \SI{138}{MHz},
\end{split}
\end{equation}
where $\epsilon = R-L/2 \simeq \SI{1}{mm}$.
This value of $\fTMS$
 implies that the frequency difference between TEM00 and TEM02 or TEM20 is \SI{276}{MHz}
 if the modes are in the same polarization.
In practice, the horizontal and vertical radii of curvature are slightly different, and TEM02 and TEM20 modes experience different frequency shifts. The separation between these frequencies can be tuned by rotating the input coupler relative to the output coupler.
The frequency difference of the two modes defines the beat note frequency used for the main readout, described in Sec. \ref{sec:layout}.
We tuned the frequency separation $\triangle f_{20/02}$ to \SI{4.5}{MHz} in order to
 minimize laser frequency noise coupling and other technical noises (see Sec. \ref{sec:cs}).

The distance from the waist to the sample mirror is $\simeq \SI{3}{mm}$,
 while to the input and output couplers are $\simeq \SI{49}{mm}$ from the waist.
These distances determine the geometry of the resonant modes in the cavity,
 and thus the beam size on the sample mirror $\omega_s \simeq \SI{55}{\mu m}$
  and on the couplers $\omega_c \simeq \SI{344}{\mu m}$. 
For our design with $\omega_{c}/\omega_{s} \simeq 7$ the CTN from the couplers
 is expected to contribute only a few percent of the total power spectral density.

\subsubsection{Optical parameters}

\def\Pin{P_{\rm in}}
\def\Pcav{P_{\rm cav}}

\begin{figure*}[t!]
\includegraphics[scale=0.5]{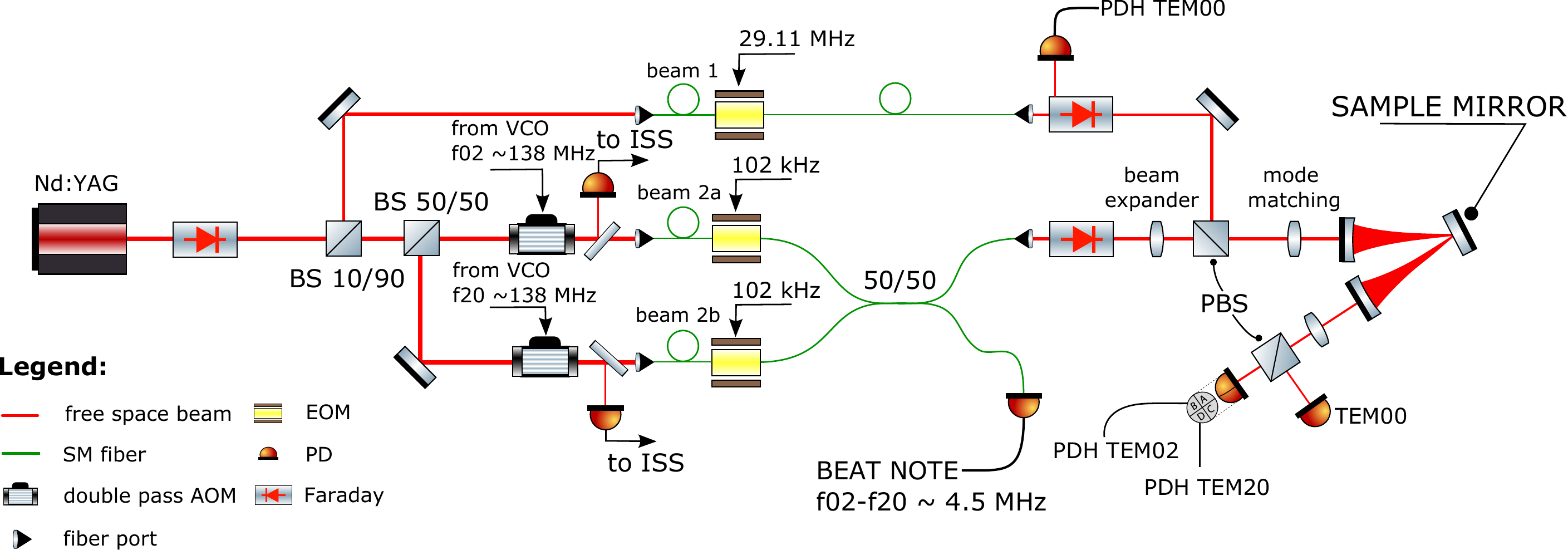}
 \caption{\label{fig:exp}The experimental setup for the multi-mode measurement involves a Nd:YAG laser (far left) and an in-vacuum high-finesse cavity (far right). In order to avoid multiple lasers, and the multiple sources of frequency and intensity noise they would introduce, a single laser beam is split into 3 paths, 2 of which are shifted in frequency (with AOMs), and each of which is independently phase modulated at its own radio frequency (with EOMs). The laser frequency is controlled to lock the TEM00 mode to the cavity length while the TEM02 and TEM20 modes are locked to the cavity by controlling their frequency differences with respect to the laser frequency. The primary output of the experiment is the difference between the TEM02 and TEM20 resonant frequencies (labeled BEAT NOTE), changes in which are dominated by the coating thermal noise of the sample.} 
\end{figure*}

The transmission of the input and output couplers was chosen to be
 $T = \SI{200}{ppm}$ (finesse of $\simeq 15000$).
The total round-trip optical loss due to the coating roughness and dust particles on the mirror surface
 was found to be $\delta \simeq \SI{20}{ppm}$.
Since our cavity is critically coupled, most of the input power is transmitted through the cavity. 

The lower limit on the beam size and the upper limit on the resonating power is determined by the following requirements.
Firstly, the intensity on the mirrors should be sustainable by the optical coating
 and be less than $\sim \SI{1}{MW/cm^2}$. 
Secondly, the beam size should be much larger than the
 coating thickness (i.e., $\omega_0 \gg d \sim 10 \, \lambda$).
The thermal propagation length in the coating,
 roughly $r_{T} \sim \SI{40}{\mu m}$ at \SI{100}{Hz} for coatings involving silica,
 is also of interest since the beams smaller than this size will experience thermo-optic noise
 which differs from those of gravitational wave detectors, where $\omega_0 \gg r_{T}$ \cite{PhysRevD.78.102003}.

For input power of $\Pin$, the power resonating in the cavity $\Pcav$ and the peak intensity $I_{c}$ on the sample mirror are approximately
\begin{equation}
\Pcav \simeq \frac{M}{T} \Pin = \SI{15}{W}
  \left(\frac{\Pin}{\SI{100}{mW}}\right)
\end{equation}
 and
\begin{equation}
I_{c} = \frac{2 \Pcav}{\pi\omega^2_{c}} \simeq \SI{380}{\frac{kW}{cm^2}} \left(\frac{ \Pin}{\SI{100}{mW}}\right)\left(\frac{\SI{50}{\mu m}}{\omega_{c}}\right)^2.
\end{equation}
For our mode matching to the cavity $M=0.03$, and beam size $\omega_{c} \simeq \SI{50}{\mu m}$, the input power of $\Pin\simeq \SI{14}{mW}$ for each higher order mode is clearly safe.

\subsection{Input and output ports}\label{sec:layout}

The optical layout is shown in Fig.~\ref{fig:exp}. The Nd:YAG laser output in the fundamental mode
 is split into three paths: 10\% of the power is fiber coupled into the TEM00 path (beam 1)
 and 45\% of the power goes into each of the TEM02 and TEM20 paths (beams 2a and 2b). 
The beams 2a and 2b are shifted in frequency relative to the beam 1 using acousto-optic modulators (AOMs).
The AOMs are configured for down-conversion, double passed, and driven at $\fTMS$
 to match the frequency of the input beam to the resonant frequencies of the cavity 02 and 20 modes.
The output beams of each AOM are coupled into optical fibers, which provide convenient transport and spatial mode stability.

All three optical paths are phase modulated using broadband electro-optic modulators (EOMs) for the feedback control presented in Sec. \ref{sec:cs}.
In addition to phase modulation with EOMs, the first and second beams are mixed using a 50/50 coupler.
The two output beams of this coupler are used to excite the cavity 02 and 20
 modes and to produce the beat note readout.
All three beams are recombined on the polarizing beam splitter (PBS) before the input to the folded cavity.

\begin{figure}[h]
\includegraphics[scale=0.6]{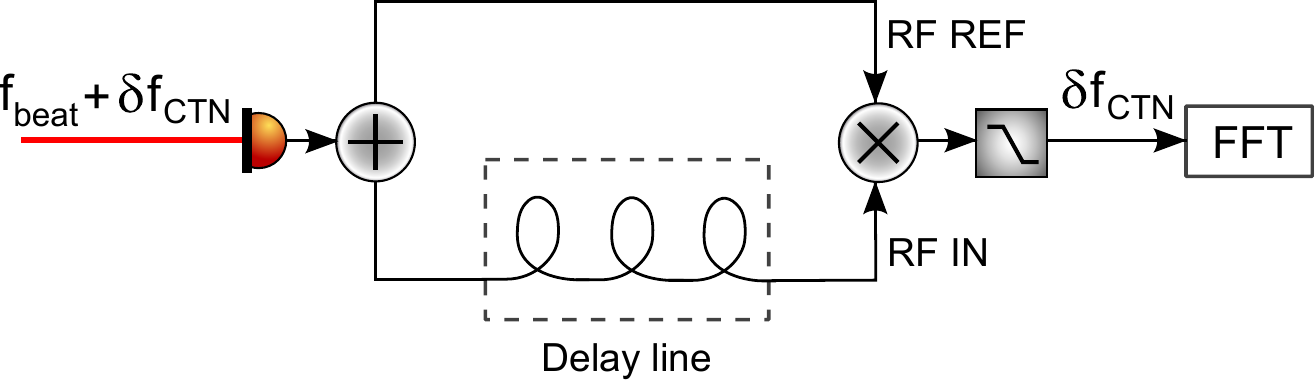}
 \caption{The delay line discriminator method for CTN readout. The conversion of frequency fluctuations imprinted on the beat note to corresponding  phase fluctuations is obtained with the delay line and subsequently converted to a voltage signal on the phase detector. Ports RF REF and RF IN add in quadrature. The $\pi/2$ phase lag between both RF ports sets the phase detector to the best linear response for the measured phase fluctuations.} 
\label{fig:rs}
\end{figure}

The spectrum of the coating thermal noise is obtained by mixing beams 2a and 2b on a photodetector.
Note that the beams 2a and 2b are in their fundamental TEM00 mode all the way to the cavity and on the beat photodetector.
The conversion to TEM02 and TEM20 transverse modes takes place in the cavity.
Optical power produces a beat note signal at \SI{4.5}{MHz},
 which is demodulated using an RF delay line, shown in Fig.~\ref{fig:rs}.
The delay line is composed of an RF splitter, a \SI{225}{m} long cable, and an RF phase detector.

\subsection{Control Scheme}\label{sec:cs}

\begin{figure}
\includegraphics[scale=0.6]{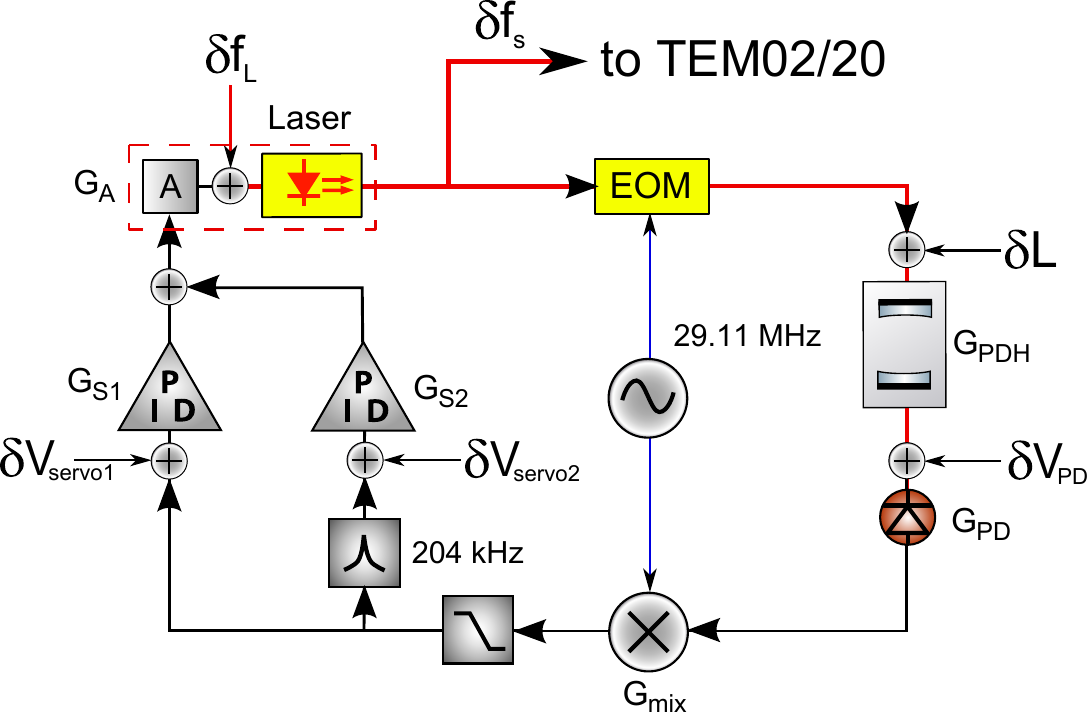}
 \caption{\label{fig:cs1}Block diagram of control loops for the TEM00 mode. All of the important noise sources are indicated as $\delta_{i}$, and each loop component with non-zero gain value is marked with $G_{i}$. The TEM00 mode is locked to the cavity length using PDH in reflection. An additional \SI{204}{kHz} loop is used to reduce frequency noise related to the locking scheme of TEM20/02 modes. More detail in text.} 
\end{figure}

Three analog servo loops are set to keep the TEM00, TEM02 and TEM20 spatial modes on resonance in the cavity.
Additional narrowband loops suppress the laser amplitude and frequency noise
 at \SI{102}{kHz} and \SI{204}{kHz} to improve sensitivity as described in Sec.~\ref{downconv}.

The laser frequency is stabilized to the cavity length using TEM00 with bandwidth of \SI{45}{kHz}. The control loop is shown in Fig.~\ref{fig:cs1}.
The Pound--Drever--Hall (PDH) error signal is derived from the reflection port.
The residual frequency noise below \SI{1}{kHz} is less than \SI{10^{-2}}{\HzrtHz}.
Ideally, the difference between TEM02 and TEM20 resonant frequencies is weakly sensitive to the residual frequency noise, however, experimental imperfections make the TEM00 loop an important
 first layer of protection from these noise sources.

\begin{figure}
\includegraphics[scale=0.6]{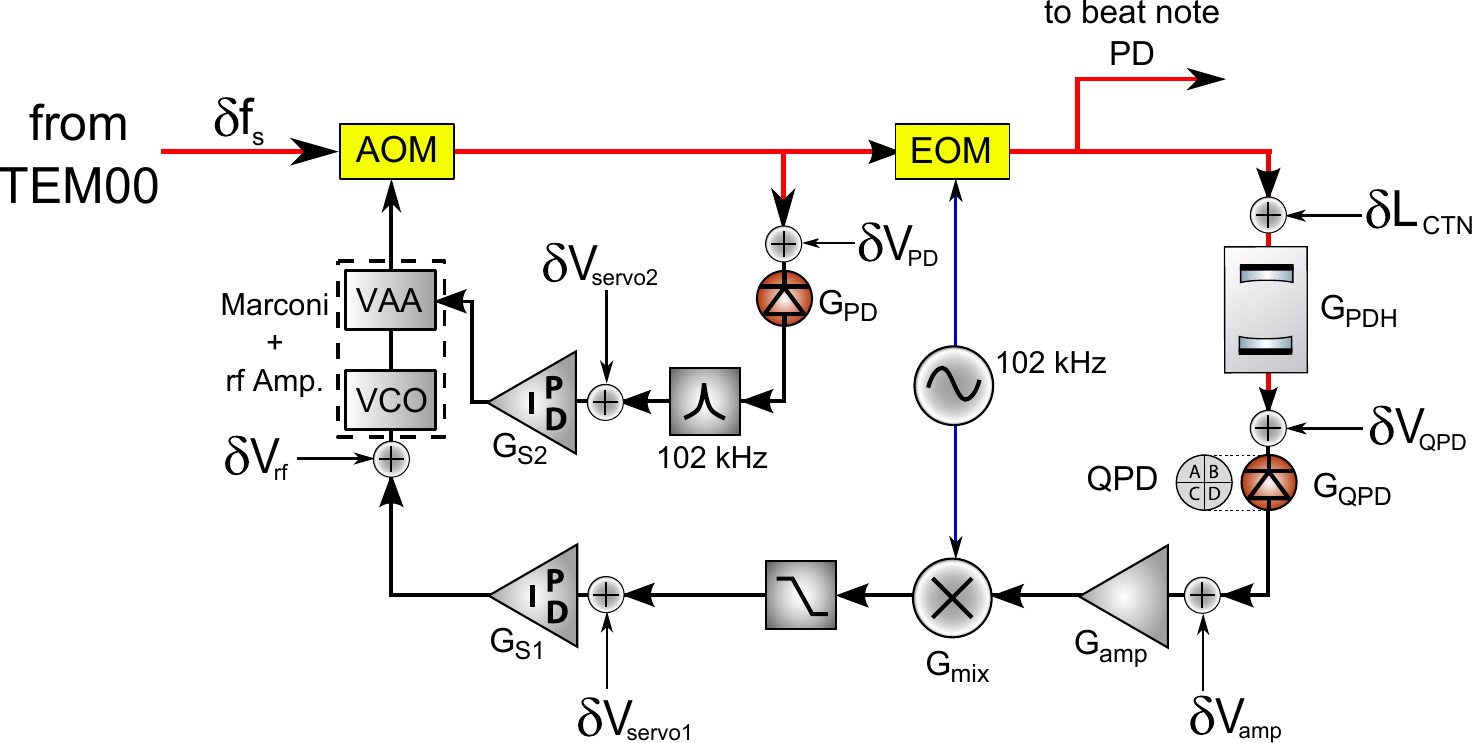}
 \caption{\label{fig:cs2} Block diagram of the TEM02/20 control loop. Both modes are frequency locked to the cavity with VCOs using PDH in transmission. An additional \SI{102}{kHz} loop is used to control intensity noise at the PDH modulation frequency.} 
\end{figure}

The frequency of RF oscillators used to shift frequencies of the beam 2a and 2b
 are stabilized to the 02 and 20 mode resonances using control loops shown in Fig.~\ref{fig:cs2}.
The error  signal is derived from the PDH signal in the transmission port.
During initial testing we found that PDH signal derived from the reflection port has extra noise
 due to the fact that only 3\% of the power from 2a and 2b beams
 is converted to TEM02 and TEM20 modes.
The residual power reflects from the cavity in the fundamental mode and adds noise to the readout.
On the other hand, only the TEM02 and TEM20 cavity modes reach the transmission port of the cavity,
 and the two modes can be separated on a quadrant photodetector.
While offering lower noise,
 this control scheme implies that modulation sidebands should also resonate in the cavity,
 and this limits their frequency $\lesssim \SI{100}{kHz}$.
This frequency limits the bandwidth of TEM02 and TEM20 modes to
 $\lesssim \SI{10}{kHz}$ and adds noises from the down-conversion
 process described in Sec.~\ref{downconv}.

\section{Extrapolation to TEM00 beams}\label{sec:C}

In this section we describe our procedure to evaluate the correction factor
 required to estimate coating thermal noise for an arbitrary TEM00 beam spot size (see Eqn. \ref{eqn:sctn1}).
Based on the value of computed here $C$,
 we estimate CTN in the Advanced LIGO gravitational wave detector, see Section~\ref{sec:aligo_ctn}.

The correction factor $C$ in Eqn.~\ref{eqn:sctn1} is defined as a product of individual factors
 related to distinct cavity parameters, 
\begin{equation}\label{eqn:sctn2}
	C = C_{\omega} \times \Cfold \times C_{\rm TEM} \times C_{\rm fringe} \times C_{d} \times C_{\rm FTM},
\end{equation} 
where $C_{\omega}$ is the ideal beam size scaling factor,
 and $\Cfold$ is the conversion factor from a folded to linear cavity.
The other correction factors, which are all close to unity, are:
 $C_{\rm TEM}$ accounts for the difference in the coating thermal noise sensed by TEM02 and TEM00 beams, $C_{d}$ corrects for the finite coating thickness, 
 and $C_{\rm fringe}$ accounts for to the fringe pattern on the sample mirror in the folded cavity\cite{PhysRevD.90.042001}.
Finally, $C_{d}$ and $C_{\rm FTM}$ correct for finite coating thickness
 and finite mirror size \cite{PhysRevD.62.122002, PhysRevD.79.102004}.

\subsection{Beam size}
In the limit of thin coatings and large optics (relative to the beam radius),
 coating thermal noise PSD simply scales inversely with area of the beam, thus
$$C_{\omega} = \left(\frac{\omega_S}{\omega_L}\right)^2,$$
where $\omega_L$ is the desired beam size (e.g., in LIGO)
and $\omega_S$ is the beam size on the sample mirror in our experiment (see Table \ref{tab:dims}).

\subsection{Higher order modes}

\begin{table}[h]
\caption{Parameters used for the calculation of correction factors. The $thick$ superscript in the correction factors corresponds to a coating thickness of \SI{6.2}{\mu m} whereas $thin$
 corresponds to the extrapolated value of for a coating which
 is much thinner than the measurement beam radius.}
\begin{center}
\begin{tabular}{lcc}
\hline
Parameter & Sample mirror & aLIGO test mass\\
\hline
Diameter & \SI{25.4}{mm} & \SI{340}{mm}\\
Thickness & \SI{6.35}{mm} & \SI{200}{mm}\\
Beam spot size & \SI{55}{\mu m} & \SI{62}{mm}\\
Pressure profile & TEM(02-20), TEM00 & TEM00\\
                 & $p_{\rm DC}$, $p_{\rm fringe}$ & \\
Substrate material & \sil & \sil \\
Coating material & \sil /\ttan:\tant &  \sil /\ttan:\tant \\
Coating model & monolayer & monolayer\\
Coating thickness, d & \SI{6.2}{\mu m} & \SI{6.2}{\mu m}\\
Analysis & harmonic & analytical\\
Frequency & \SI{100}{Hz} & DC\\
Element type &  solid185 & N/A\\
Output &$E^{\rm thick}_{HOM}$, $E^{\rm thick}_{00}$,  $E^{\rm thin}_{00}$ & $E_{\rm FTM}$, $E_{\rm INF}$\\
       &$E^{\rm thick}_{\rm DC}$, $E^{\rm thick}_{\rm fringe}$ & \\
 \hline
\end{tabular}
\end{center}
\label{tab:dims}
\end{table}

The correction factors $C_{\rm TEM}$,  $C_{d}$, and $C_{\rm fringe}$ are calculated using the stored strain energy in the coating.
Since dissipation is a product of the stored energy and the tangent
 of the coating material loss angle,
 the spectral density of the coating thermal noise scales with energy \cite{0034-4885-29-1-306}.

\begin{figure} [t]
\includegraphics[scale=0.25]{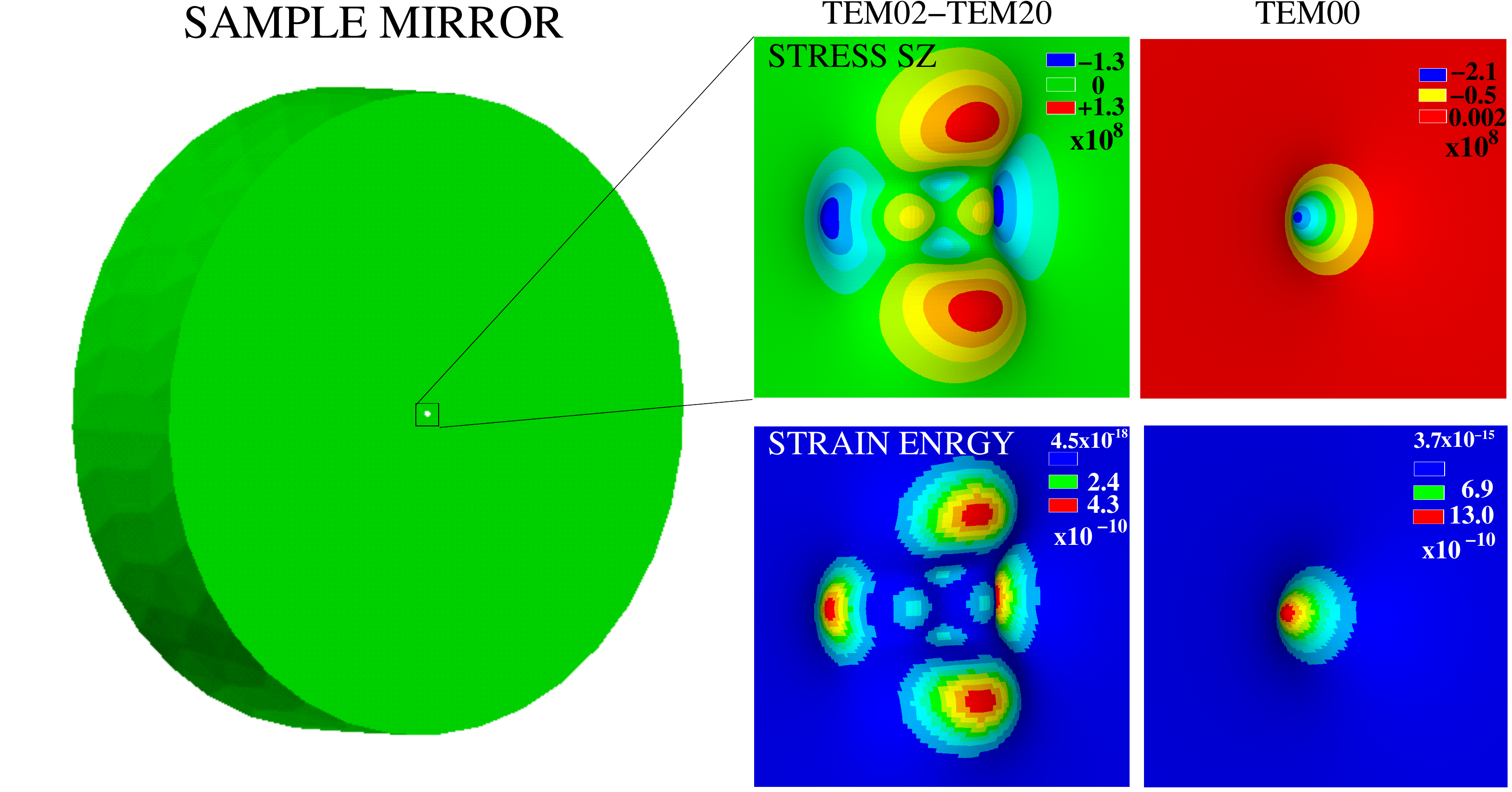}
 \caption{Stored energy distribution and stress concentration as a result of an applied pressure field on the coating surface. For reference we also show TEM00 results. It is worth noting that the differential intensity profile corresponding to TEM20/02 causes zero net-stress at the rim of the sample mirror. Thus this multi-mode CTN measurement technique is insensitive to the losses associated with mirror clamps.}
\label{fig:mesh1}
\end{figure}

The strain energy associated with the TEM00 mode, $E_{00}$,
 is produced by a pressure profile associated with the optical field intensity, $\Psi$, \cite{PhysRevD.57.659}
$$p_{00} = \int\Psi_{00}\Psi_{00}^{*}d\vec{r}$$
 while for the experiment readout, the energy, $E_{02/20}$,
 results from the pressure profile
$$p_{02/20} = \int\Psi_{02}\Psi_{02}^{*}d\vec{r} - \int\Psi_{20}\Psi_{20}^{*}d\vec{r}$$
The negative sign between TEM02 and TEM20 intensities corresponds to their opposite sign in the readout, and results in the signal cancellation
 of the overlapping central part of the TEM20 and TEM02 modes, see Figs.~\ref{fig:cavity} and \ref{fig:mesh1}.

In order to obtain strain energy corresponding to each pressure profile,
 we performed a harmonic finite element analysis~\cite{ansys} 
 in which an oscillating pressure profile with dimensions and shape of the
 optical mode is applied onto the coating surface. 
Both the substrate and coating models were meshed with low order 3D elements
 and the coating was simulated as a monolayer with effective material properties~\cite{PhysRevD.91.022005}.
An example of calculated energy profiles in the coating is shown in FIG.~\ref{fig:mesh1}.

We define the multi-mode correction factor as
\begin{equation}
 C_{\rm TEM} = \frac{E^{\rm thick}_{00}}{E^{\rm thick}_{\rm HOM}}=1.13,
\end{equation}
 where $E_x$ is the strain energy stored in a coating corresponding to the TEM00 and TEM02-20 modes,
 and the superscript ``thick'' indicates that this energy is computed with a finite coating thickness, see Table~\ref{tab:dims}.
The value of  $C_{\rm TEM}$ indicates that TEM02 and TEM20 sense a smaller coating thermal noise compared to the TEM00 of the same Hermite-Gaussian modal basis.
 Interestingly, $C_{\rm TEM}$ goes to unity for a thin coating on a stiff substrate, and thus is entirely due to the details of the mechanical response
 of the mirror to the applied pressure profile.

\subsection{Folded cavities}

The folded cavity conversion factor $\Cfold$ is the ratio of the coating thermal noise of the folded cavity to that of a linear cavity,  
$$\Cfold = \frac{S_{\rm linear}}{S_{\rm folded}}=\frac{1}{4},$$
 since the sample mirror is encountered twice in a cavity round-trip.
This doubles the \emph{amplitude} of the coating thermal noise, and thus requires a factor of 4 correction in the power spectrum density.

\begin{figure}
\includegraphics[scale=0.20]{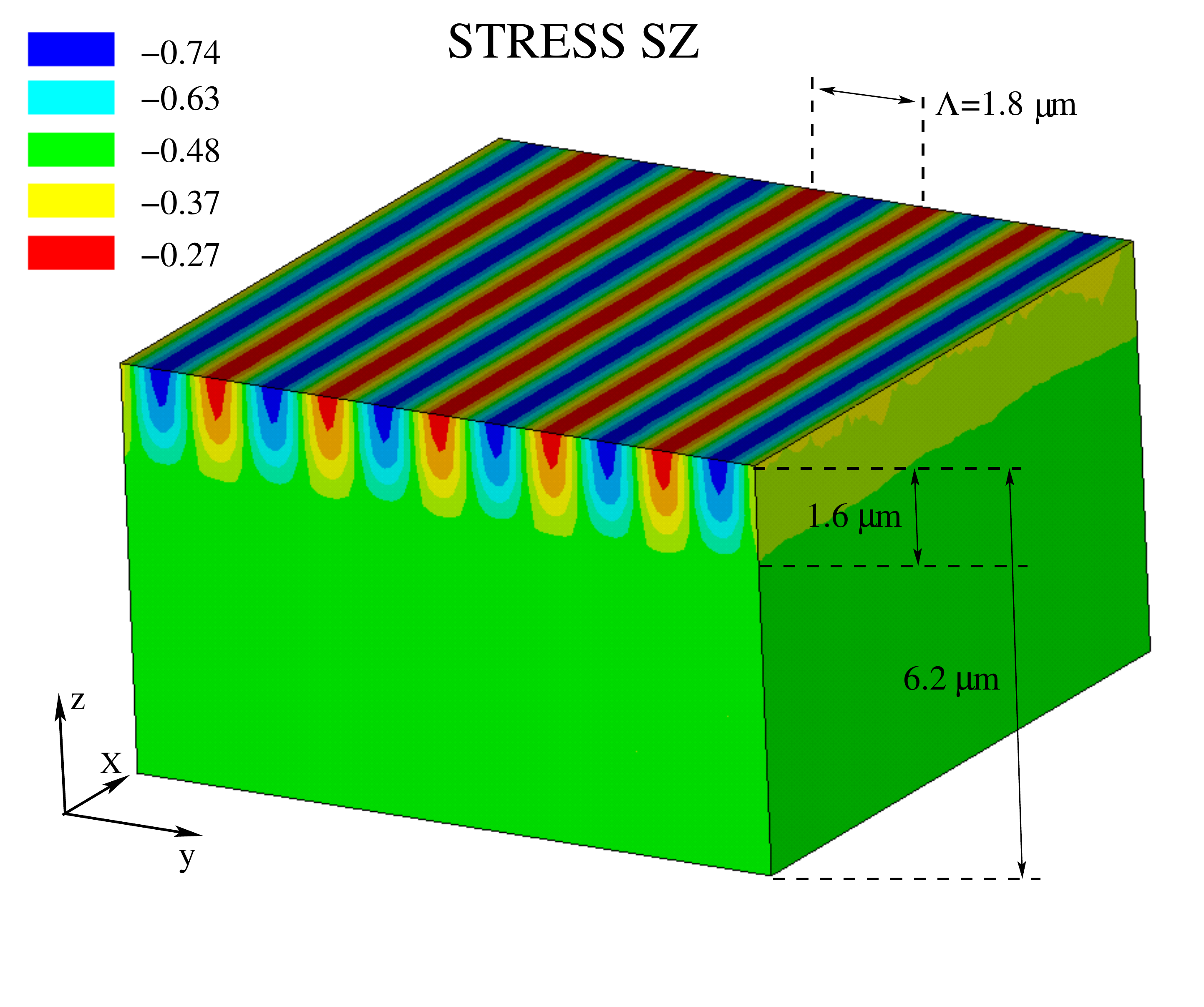}
 \caption{Counter-propagating beams on the folding mirror cause an interference pattern imprinted on the resonating modes. The fringe pattern can affect the sensitivity to coating thermal noise. For a cavity folding angle of 17.2 deg (Table \ref{tab_cav}) the fringe separation is \SI{1.8}{\mu m}. The figure shows the stress in the coating in the z-direction as a result of an applied pressure profile $p_{\rm fringe}$. The coating is only affected by the fringe down to the depth of \SI{1.6}{\mu m} below which the stress becomes uniform.
For our folded cavity the correction factor $C_{\rm fringe} = 0.98$ is only slightly different from unity,
 indicating that this effect has little impact on our results.} 
\label{fig:fringe}
\end{figure}

The fringe correction factor $C_{\rm fringe}$ is related to the fact that folded cavities have increased thermal noise due to the fringe pattern on the folding mirror \cite{PhysRevD.90.042001}.
For the folding angle of $17.23^{\rm o}$ used in our testbed
\begin{equation}
 C_{\rm fringe} = \frac{E_{\rm DC}^{\rm thick}}{E_{\rm fringe}^{\rm thick}}=0.98,
\end{equation}
 where $E_{\rm DC}$ corresponds to the energy due to the uniform pressure field
 $p_{\rm DC} =1/2$ applied to the mirror and $E_{\rm fringe}$
 is the energy corresponding to the pressure profile
 $$p_{\rm fringe}= \cos^2 \left(\frac{\pi}{\Lambda}\cdot y \right)$$
 with fringe separation $\Lambda=\SI{1.79}{\mu m}$, see Fig. \ref{fig:fringe}.
Note that this is rather different than the limit of an infinitesimally thin coating
 in which $C_{\rm fringe}$ approaches $2/3$.

\begin{figure*}[t]
\centering
\includegraphics[scale=0.4]{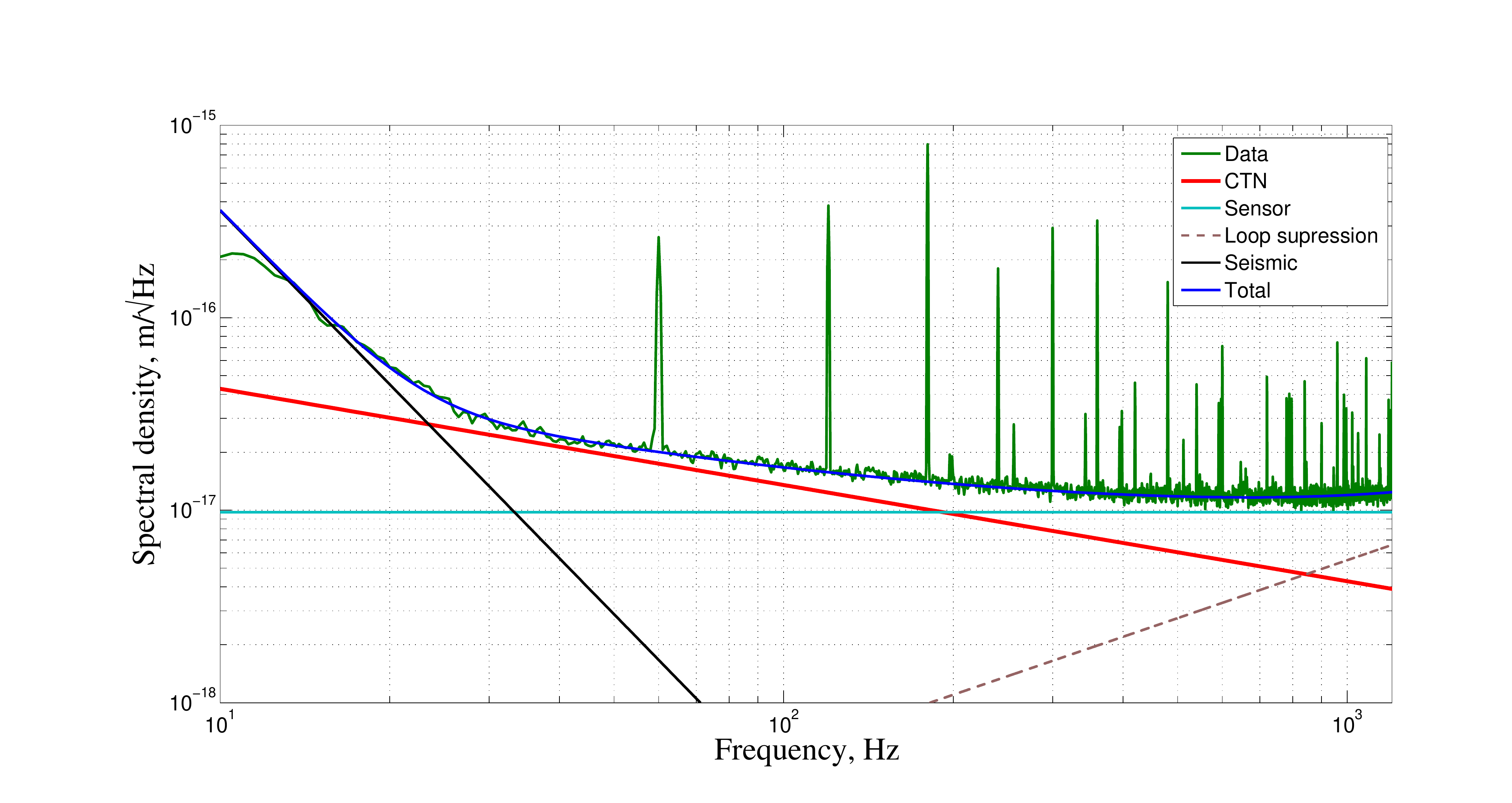}
 \caption{The measurement result for an Advanced LIGO ETM coating witness sample.
 In the range \SI{30-400}{Hz} the largest noise contributions comes from coating thermal noise
  with a characteristic $1/\sqrt{f}$ slope, and a white sensing noise.
 Most of the sharp features are from the \SI{60}{Hz} mains.} 
\label{fig:mp}
\end{figure*}

\subsection{Finite coating size}

The thick to thin coating thickness correction factor is defined as 
\begin{equation}
 C_{d} = \frac{E_{00}^{\rm thin}\cdot d_{\rm thick}}{E_{00}^{\rm thick}\cdot d_{\rm thin}}=1.33,
\end{equation}
where $d_{\rm thick}$ corresponds to the actual coating thickness and $d_{\rm thin} \ll d_{\rm thick}$
 is chosen to approximate a coating which is much smaller than the target beam size,
 assuming $d_{\rm thick} \ll \omega_L$.
The result indicates that the strain energy density in the thick coating does not fully reach the bottom layers, and is thus slightly smaller than in the thinner coating.

The finite mirror size correction factor $C_{\rm FTM}$ is not directly related to the coating thermal noise measurement,
 but is required to correctly compute this noise in Advanced LIGO \cite{PhysRevD.62.122002, PhysRevD.79.102004}.
The finite size correction factor for a large beam spot size is defined as
\begin{equation}
 C_{\rm FTM} = \frac{E_{\rm FTM}}{E_{\rm INF}}=1.03,
\end{equation}
where $E_{\rm FTM}$ corresponds to the coating energy of the finite size mirror to which the coating thermal noise measurement is extrapolated and $E_{\rm INF}$ corresponds to the energy of an infinite mirror.
To calculate $C_{\rm FTM}$ we used equations 26, 27, and 28 from \cite{PhysRevD.79.102004}.

\section{Experimental Results}\label{res}

\begin{table}[b!!!]
\caption{Measured cavity parameters during collection of the data.}
\begin{center}
\begin{tabular}{lcc}
\hline
Parameter & TEM02 & TEM20\\
\hline
 Input power, mW & 14.2 & 14.3\\
Transmitted power, mW & 0.4 & 0.4\\
Cavity pole, kHz & 50.8 & 50.0\\
Modulation, kHz & 102 & 102\\
Modulation depth & 0.92 & 0.92\\
 Finesse, $10^3$ & 15.06 & 15.30	\\
Round trip loss, ppm & 17.3 & 10.7\\
Mode coupling, \% & 3.1 & 3.0\\
Mode frequency, MHz &  276.462 & 280.914 \\
Beam size, $\mu$m  & 55.10 & 55.38\\
RoC (effective), mm & 50.883 & 50.919\\
Arm length, mm &\multicolumn{2}{l}{ $L_1+L_2=46.45+53.07$} \\
Folding angle, deg & \multicolumn{2}{c}{17.23}\\
 \hline
\end{tabular}
\end{center}
\label{tab_cav}
\end{table}

This section describes the sensitivity of the experimental setup and results for the Advanced LIGO coating sample~\cite{0264-9381-32-7-074001}.
The tested coating, produced at Laboratoire des Mat\'{e}riaux Avanc\'{e}s (LMA, Lyon, France),
 was ion-beam sputtered and consists of alternating layers of
 SiO$_2$ and Ta$_2$O$_5$ alloyed with 25\% TiO$_2$. 
The thickness of \sil\ and \ttan:\tant\ layers was optimized to operate at
 \SI{1064}{nm} and \SI{532}{nm}
 (for more details on coating structure, see Table VII in \cite{PhysRevD.87.082001}).
The sample mirror has a transmissivity of $T = \SI{5}{ppm}$ at the measurement
 wavelength ($\lambda = \SI{1064}{nm}$).

The measured amplitude spectral density is shown in Fig. \ref{fig:mp}
 and the folded cavity parameters are shown in Table~\ref{tab_cav}.
The slope in the frequency range \SI{30-400}{Hz} is a combination of the
 $1/\sqrt{f}$ coating thermal noise $N_{\rm CTN}$ and white sensing noise.
This noise is a sum of multiple contributors, described in the following section.

\subsection{Noise sources}

The measured noise PSD $N_{02/20}$ contains the coating thermal noise
 $N_{\rm CTN}$ as well as fundamental and technical noises.
In this section we describe the major noise sources that contribute to $N_{02/20}$.
The only ``fundamental'' limit to sensitivity is the shot noise on the readout,
 but technical noise sources such as photo-thermal noise,
 vibration coupling through scattered light, and RF oscillator noise are also discussed
 here since they have the potential to limit the sensitivity of this technique.

\subsubsection{Vibration Noise}\label{res:vibration}

\def\Nlen{N_{\rm len}}

Mechanical motion of the optical table couples to the readout channel via several mechanisms and limits the measurement below \SI{20}{Hz}, as seen in Fig. \ref{fig:mp}.

First of all, fluctuations of the cavity total length $\Nlen$ couple to the readout channel very weakly due to the common mode rejection.
Residual coupling is due to the frequency difference $\triangle f_{02/20}$ between 02 and 20 modes:
\begin{equation}
	N_{02/20} = \frac{\triangle f_{02/20}}{f_0} ~\Nlen 
	= 1.4 \times 10^{-8} \Nlen,
\end{equation}
where $f_0 = 2.82\times 10^{14}$\,Hz is the laser frequency.
The measurement of the cavity length fluctuations is limited by the laser frequency noise
 $$\Nlen(f) \simeq 2 \times 10^{-13} \left( \frac{\SI{10}{Hz}}{f} \right) \mrtHz.$$
Coupling to the readout channel at \SI{10}{Hz} is $\simeq 3 \times10^{-21} \mrtHz$,
 which is four orders of magnitude below CTN.

Vibrations of the cavity mirrors can also couple to the readout channel through backscattering.
Since only the total length of the cavity is controlled, optical phase $\theta$
 between the couplers and the sample mirror is uncontrolled.
Fluctuations of this phase introduce extra noise to the readout channel is
\begin{equation}
N_{02/20}(f) = (\alpha_{02} \cos\theta_{02} - \alpha_{20} \cos\theta_{20}) N_{L_1}(f), 
\end{equation}
where $N_{L_1}(f)$ is vibration of the distance between the input coupler and the folded mirror at frequency $f$.
The backscattering coefficient $\alpha$ is determined by the equation
\begin{equation}
	\alpha = \sqrt{\text{BRDF} \frac{\lambda^2}{\pi \omega_0^2}} \sim 10^{-5},
\end{equation}
where $\text{BRDF} \sim 10^{-6} \text{\,sr}^{-1}$ is the the bidirectional reflectance distribution function of the sample mirror.
Backscattering coefficients $\alpha_{02}$ and $\alpha_{20}$
 can be slightly different since TEM02 and TEM20 reflect from the different parts of the mirror surface.
 Backscattering adds noise to the readout channel on the order of $N_{02/20} \sim 10^{-5} N_{L_1}$.
We estimate this as $2\times10^{-18} \text{\,m} / \sqrt{\text{Hz}}$ which is still an order of magnitude below the coating thermal noise.
However, backscattering also occurs outside of the cavity and adds noise to the readout channel.

Finally, mechanical motion of the input mirrors also introduces noise to the readout channel in two ways. First, longitudinal motion $L_{\rm input}$ modulates the phase of the beams 2a and 2b according to the equation
\begin{equation}
N_{20/02}(f) = \frac{L_{\rm input}}{\lambda} 2\pi \frac{f}{f_0} L.
\end{equation}
Our estimations show that $L_{\rm input} \sim \SI{10^{-9}}{\mrtHz}$ at \SI{10}{Hz}.
This noise couples to the readout channel at this frequency at the level of $N_{20/02} \sim \SI{10^{-17}}{\mrtHz}$.
This number is close to the coating thermal noise at \SI{10}{Hz}.
Secondly, angular motion of the input mirrors modulates the power resonating in the cavity. Power fluctuations couple to the readout channel through the photo--thermal noise.

\subsubsection{Photo-thermal Noise}

\def\Spt{S_{\rm PT}}
\def\Srin{S_{\rm RIN}}

Power fluctuation in the cavity can couple to cavity length through thermal expansion of the sample mirror, resulting in a length noise known as ``photo-thermal'' or sometimes ``thermo-optic'' noise (not to be confused with the coherent combination of thermo-elastic and thermo-refractive noise also known  as ``thermo-optic'' noise \cite{PhysRevD.78.102003}).
  
The expression for the thermal expansion in the case where the beam size is large compared to the thermal propagation length in the substrate, $\tau_{T}=\sqrt{\kappa/2\pi C_{\rho}}$, is
\begin{equation}
N_{02/20}^2(f|\omega_{c} \gg r_{T})=\frac{\Pabs \alpha (1+\sigma)\Srin}{2\pi^2 f C_{\rho} \omega^2_c},
\end{equation}
where $\Pabs$ is the absorbed power, $C_{\rho}$ is the heat capacity per unit volume (written as the product of the heat capacity per unit mass and the density), $\alpha$ is the coefficient of thermal expansion, $\Srin$ is the power spectrum density of the laser relative intensity noise, and $\sigma$ is the Poisson ratio (see Sec. 2.8.5 in \cite{ballmerthesis}). In the opposing limit of low frequencies $\omega_{c}$ is replaced by $r_{T}$ \cite{PhysRevA.44.7022}.
\begin{equation}
N_{02/20}^2(f|\omega_{c} \ll r_{T})=\frac{\Pabs \alpha (1+\sigma)\Srin}{8\pi^2 f C_{\rho} r^2_{T}},
\end{equation}

A simple combination of these is
\begin{eqnarray}
\nonumber
N_{02/20}^2 &\simeq& \frac{\Pabs \alpha (1+\sigma)\Srin}{2\pi^2 f C_{\rho} \sqrt{(2r_{T})^4+\omega_{c}^4}}\\
\nonumber
&\simeq&  10^{-19} \frmrtHz \frac{\Pabs}{\SI{10}{\mu W}}
  \frac{\alpha(1+\sigma)}{\SI{10^{-6}}{/K}} \\
  &&\times \frac{\Srin}{\SI{10^{-7}}{/\rtHz}} 
  \frac{\SI{100}{Hz}}{f}\frac{\SI{164}{kJ/Km^3}}{C_{\rho}}
\end{eqnarray}
 which agrees with an experimental limit we placed on this coupling of less than \SI{4 \!\times\! 10^{-19}}{\mrtHz}
  for a RIN of \SI{10^{-6}}{/\rtHz}.

A similar but more subtle noise source is the change in curvature of the optic resulting from heating of the coating, which in turn changes the transverse mode spacing and could appear in the primary output.  The calculation of this ``thermo-optic curvature noise'' will not be reproduced here, since the result is numerically smaller than the direct length coupling by more than an order of magnitude.

\subsubsection{RF Oscillator Noise}\label{lsi}

RF oscillators, used to shift the frequency of higher order modes (see Fig.~\ref{fig:cs2}), have noise level on the order of $N_{\text{osc}} \sim 10^{-1}\HzrtHz$ below 1\,kHz.
This noise is suppressed by the feedback loops which keep TEM02 and TEM20 on resonance.
However, finite bandwidth of these loops results in the limited suppression $G(f)$ of the RF oscillator noise,
 which then adds noise to the readout channel according to the equation
\begin{equation}
	N_{02/20}(f) = \frac{1}{G(f)} \frac{N_{\text{osc}}}{f_0} L \sim \frac{1}{G(f)} 10^{-17} \frac{\text{m}}{\sqrt{\text{Hz}}}.
\end{equation}
RF oscillator noise causes the degradation of the sensitivity above \SI{400}{Hz}. The noise floor rises towards the unity gain frequency of higher order feedback control loops. (see Fig. \ref{fig:mp}).

\subsubsection{Readout electronics}

In order to prevent any environmental RF pickup and seismic noise, the delay line (see Sec.~\ref{sec:layout}) is enclosed in a thick metal chamber and wire suspended.
The noise related to the readout system (RF Oscillator + Delay line) is estimated at the level of
$\simeq~\SI{2.0 \times 10^{-3}}{\HzrtHz}$.

Converting this to the units of $\mrtHz$ of the cavity length,
$$
N_{02/20} = \SI{6.8 \times 10^{-19}}{\mrtHz}.
$$
which is a factor of $\simeq$ 20 below the level of the coating thermal noise at  100\,Hz.

\subsubsection{Shot noise}

\def\Sshot{S_{\rm shot}}

Photon counting noise, or ``shot noise'', is an unavoidable noise source in precision optical measurements
 and sufficient power on the sensor is required to sustain the shot noise below the coating thermal noise level.
The relevant equation for shot noise in the PDH readout of a high-finesse cavity, expressed as an equivalent displacement of the optics is
\begin{equation}
N_{02/20} = \frac{\lambda}{8 F}\sqrt{\frac{2h\nu}{\Pin}}
 \frac{\sqrt{1-J^2_{0}(\beta)M}}{\sqrt{2}MJ_{0}(\beta)J_{1}(\beta)},
\end{equation}
where $J_{n}(\beta)$ are the Bessel functions, and $\lambda = \SI{1064}{nm}$ is the wavelength of light used in the cavity \cite{Fritschel}.
To compute a shot noise level it is further assumed that the mode matching is $M = 0.03$ and modulation depth is $\beta = \SI{0.8}{rad}$. The resulting shot noise, assuming a total input power of \SI{3}{mW} and a finesse of $F = 1.5 \times 10^{4}$, is
\begin{equation}
N_{02/20} \simeq \SI{7.6 \times 10^{-18}}{\mrtHz}
\end{equation}

\subsubsection{Down-conversion}\label{downconv}

A significant fraction of the observed broadband white noise at the level of $10^{-17} \text{\,m}\sqrt{\text{Hz}}$
 can be explained by the process of downconversion.
High frequency laser amplitude and frequency noises are seen in the audio band due to
 the non-linear demodulation processes required to produce PDH error signals.

Differential amplitude fluctuations of TEM02 and TEM20 beams at the modulation frequency (\SI{102}{kHz})
 directly couple to the readout channel.
These fluctuations arise from the imbalance in the cavity poles for two modes and due to different
 input paths of the beams 2a and 2b.
This noise was suppressed by using the additional intensity stabilization servos described in Sec.~\ref{sec:cs},
 without which it would be a factor of 3 above shot noise.
The readout channel sees down-converted amplitude noise at the level of
\begin{equation}
	N_{02/20} = 2 \times 10^{-17}\frac{1}{G_{\text{iss}}}{\mrtHz},
\end{equation}
where $G_{\text{iss}} \simeq 6$ is the open loop gain of the intensity stabilization servo around 102\,kHz.

Secondly, frequency noise around 102\,kHz and harmonics is down-converted to the audio band during the demodulation process. We found that the biggest contribution comes from the noise around the second harmonic at 204\,kHz. An addition servo has been introduced to suppress laser noise around this frequency as shown in the Fig.~\ref{fig:cs1}.

Finally, RF frequency noise around the beat frequency $\triangle f_{02/20}$ is downconverted to the audio band if there is an imbalance in the pole frequencies for TEM02 and TEM20 modes. A careful analysis of optical losses and cleaning the mirrors helped to reduce this imbalance as shown in the Table~\ref{tab_cav}.
The beat frequency $\triangle f_{02/20}$ was also set to minimize the laser noise at this frequency.

\subsection{Advanced LIGO Coating Thermal Noise}\label{sec:aligo_ctn}

The least square fitting of a series of spectra gives the following result for the coating thermal noise in our experiment:
$$N_{\rm CTN} = {(1.29\pm 0.06)\times10^{-17}} \sqrt{\frac{\SI{100}{Hz}}{f}}\sqrt{\frac{\rm{T}}{\SI{300}{K}}} \frmrtHz. $$

Extrapolation of our measured PSD to the PSD of a large beam on an aLIGO end test mass,
 the total correction factor can be written as
\begin{equation}
 C = \left(\frac{\omega_S}{2 \, \omega_L}\right)^2 \times 1.52 = 2.99\times 10^{-7}
\end{equation}

Based on Eqn.~\ref{eqn:sctn1}
 we estimate the value of the coating thermal noise for the Advanced LIGO end test mass (ETM):
\begin{eqnarray}
 N_{\rm CTN}^{\rm 00} &=& \sqrt{C} \times
 (1.29\pm0.06)\times10^{-17} \nonumber \\
&=&(7.1 \pm 0.3)\times10^{-21} \sqrt{\frac{\SI{100}{Hz}}{f}}\sqrt{\frac{\rm {T}}{\SI{300}{K}}}\frmrtHz
\end{eqnarray}
which is slightly higher than the value used in Advanced LIGO design documents ($5.9 \times10^{-21}\sqrt{\frac{\rm {T}}{\SI{300}{K}}}~\mrtHz$ at \SI{100}{Hz} calculated with Eqn.1,2 in \cite{0264-9381-24-2-008} and for the loss angle value of $\phi_{\rm Si02}=4.0\times10^{-5}$ and $\phi_{\rm Ti:Ta}=2.3\times10^{-4}$ \cite{0264-9381-32-7-074001},).\\
Since the Advanced LIGO input test mass coating is made out of the same materials, we estimate an overall increase of the coating thermal noise by 20\% compared to \cite{0264-9381-32-7-074001}. Interestingly, there is some evidence that this higher estimation of the coating thermal noise can be associated with interface losses in the coating structure as reported in \cite{PhysRevD.93.012007}. 

\subsection{Loss angle of \ttan:\tant} 
  
To estimate the loss angle for the titania-tantala alloy used as the high refractive index material
 in the Advanced LIGO coatings, we use the loss angle for
 silicon-dioxide (the low index material) of $\phi_{\rm Si02} = 5\times10^{-5}$ \cite{PhysRevD.91.022005},
  and assume that the loss angles associated with shear and bulk deformation in
  both materials are equal.

We adopted the formula from \cite{PhysRevD.91.042002} and calculate the power spectrum density
\begin{equation}
S = \frac{2k_{B}T}{\pi^2 f \omega^2_{c}}\frac{1-\sigma_{s}-2\sigma^2_{s}}{Y_{s}}\sum_{j}b_{j}d_{j}\phi^M_{j}
\end{equation} 
where the unitless weighting factor $b_{j}$ for each layer is
\begin{equation}
b_{j} = \frac{1}{1-\sigma_{j}}\left[\frac{Y_{s}}{Y_{j}}+\frac{(1-\sigma_{s}-2\sigma^2_{s})^2}{(1+\sigma_{j})^2(1-2\sigma_{j})}\frac{Y_{j}}{Y_{s}}\right],
\end{equation}
under the approximation that no field penetrates into the coating.

Our estimation for the loss angle is 
 $\phi_{\rm Ti:Ta}={(3.1\pm 0.5})\times 10^{-4}$. This number is slightly lower than the value previously reported in \cite{PhysRevD.91.022005}, but higher than the value reported in \cite{0264-9381-27-8-084030}.

\vspace{20pt}
\section{Conclusions}

We presented a novel experiment for the broadband direct measurements of the coating thermal noise.
The sensitivity of \SI{10^{-17}}{\mrtHz} has been achieved in the frequency band \SI{30 - 1000}{Hz}.
This is made possible by our novel measurement technique,
 in which TEM00, 02 and 20 spatial modes all co-resonate in a folded cavity.

As a first application of this technique,
 we measured the coating thermal noise from Advanced LIGO coating and estimated the loss angle of \ttan:\tant. 
Our results are broadly consistent with the previous estimations, but give a 20\% higher coating thermal noise compared to the published Advanced LIGO noise estimates~\cite{0264-9381-32-7-074001, den_nb}.

With the ever increasing sensitivity of precision optical measurements,
 coating thermal noise has become a significant obstacle.
In terms of the gravitational wave interferometers and some macroscopic quantum measurement experiments,
 it is essential to reduce this noise in order to reach and surpass the standard quantum limit.
Our experiment design will allow for rapid testing new coatings,
 thereby helping to reduce the coating thermal noise
 in the future generation of gravitational wave detectors, frequency references and quantum measurements.

\begin{acknowledgments}
The authors would like to acknowledge the invaluable wisdom derived from
 interactions with members of the LIGO Scientific Collaboration's optics working group
 without which this work would not have been possible.
In particular, the multi-mode cavity design developed from a seed planted several years ago in a conversation with Koji Arai.
We are also very grateful for the computing support provided by The MathWorks, Inc.

LIGO was constructed by the California Institute of Technology and
 Massachusetts Institute of Technology with funding from the National Science Foundation,
 and operates under cooperative agreement PHY-0757058.
Advanced LIGO was built under award PHY-0823459.
This paper carries LIGO Document Number LIGO-P1600228.
\end{acknowledgments}

\bibliography{aLIGO_CTN}

\end{document}